\documentclass[prd,aps,twocolumn,amsmath,amssymb,nofootinbib,preprintnumbers]
{revtex4}

\usepackage{graphicx}
\usepackage{dcolumn}
\usepackage{bm}
\usepackage{amsmath}
\usepackage{amsfonts}

\newcommand{\be}{\begin{equation}}
\newcommand{\ee}{\end{equation}}
\newcommand{\bea}{\begin{eqnarray}}
\newcommand{\eea}{\end{eqnarray}}
\newcommand{\ba}{\begin{eqnarray}}
\newcommand{\ea}{\end{eqnarray}}

\newcommand{\beq}{\begin{equation}}
\newcommand{\eeq}{\end{equation}}
\newcommand{\beqa}{\begin{eqnarray}}
\newcommand{\eeqa}{\end{eqnarray}}
\newcommand{\beqar}{\begin{eqnarray*}}
\newcommand{\eeqar}{\end{eqnarray*}}
\newcommand{\e}{\epsilon}

\newcommand{\la}{\lambda}

\def\gev{\; \rm{GeV}}
\def\mev{\; \rm{MeV}}
\def\De{\Delta}
\def\a{\alpha}
\def\b{\beta}

\def\te{\tau_\Pi^\eta}
\def\tz{{\tau_\Pi^\zeta}}
\def\N{\mathcal{N}}
\def\MeV{\; \rm{MeV}}

\def\gsim{\mathrel{\rlap{\lower4pt\hbox{\hskip1pt$\sim$}}
    \raise1pt\hbox{$>$}}}

\newcommand{\grqc}[1]{gr-qc/#1}

\newcommand{\pla}[3]{Phys. Lett. B {\bf #1}, #3 (#2)}

\renewcommand{\ap}[3]{Ann. Phys. (N.Y.) {\bf #1}, #3 (#2)}

\renewcommand{\prd}[3]{Phys. Rev. D {\bf #1}, #3 (#2)}
\newcommand{\jmp}[3]{J. Math. Phys. {\bf #1}, #3 (#2)}

\begin{document}

\title{Cavitation in holographic sQGP}

\author{Aleksandra Klimek$^1$}
\author{Louis Leblond$^2$}
\author{Aninda Sinha$^3$}

\affiliation{$^1$ Department of Physics, University of Warsaw, Poland\\
$^2$Perimeter Institute for Theoretical Physics, Waterloo, Ontario N2L 2Y5, Canada\\
$^3$ Centre for High Energy Physics, Indian Institute of Science,
C. V. Raman Avenue, Bangalore 560012, India}

\begin{abstract}
We study the possibility of cavitation in the nonconformal ${\mathcal N}=2^*$ $SU(N)$ theory which is a mass deformation of ${\mathcal N}=4$ $SU(N)$ Yang-Mills theory. The second order transport coefficients are known from the numerical work using AdS/CFT by Buchel and collaborators. Using these and the approach of Rajagopal and Tripuraneni, we investigate the flow equations in a 1+1 dimensional boost invariant set up. We find that the string theory model does not exhibit cavitation before phase transition is reached. We give a semi-analytic explanation of this finding.
\end{abstract}
\preprint{pi-strings-215}
\maketitle
\section{Introduction}

Viscous relativistic hydrodynamics has been applied successfully to the strongly coupled Quark Gluon Plasma (sQGP) created in nuclear collisions at Relativistic Heavy Ions Collider (RHIC). The fact that a low shear viscosity to entropy density ratio seems to be favoured from this experiment  coupled with the fact that any holographic transport calculation using AdS/CFT also gives rise to a low value for this ratio \cite{Policastro:2001yc,Kovtun:2004de} has triggered a huge amount of activity in using AdS/CFT methods to understand cousins of the sQGP (see \cite{CasalderreySolana:2011us} for a recent review).  Since then, other hydrodynamic coefficients have been obtained
using AdS/CFT. Of particular interest for this work is the bulk viscosity $\zeta$ and the relaxation times $\tau_\pi^\eta, \tau_\Pi^\zeta$. 

While the bulk viscosity is usually very small in sQGP, there is good evidence that it rises sharply near the critical temperature \cite{Paech:2006st,Kharzeev:2007wb,Meyer:2007dy,Karsch:2007jc, Sasaki:2008fg} (see also  \cite{Lu:2011df, Bluhm:2010qf}).
With high enough viscosity, the boost-invariant solution to the hydrodynamics equations become unstable against small perturbations \cite{torri}.
A related effect arises due to the fact that the bulk viscosity affects the trace of the stress-energy tensor and it can lower the effective longitudinal pressure. 
The longitudinal pressure of the plasma could become negative making the plasma unstable to clustering and cavitation. The proposition is then that this cavitation process could be one possible description of hadronization of the sQGP. 

Cavitation occurs when a fluid has such a large gradient velocity and bulk/shear viscosity coefficient that the viscous correction to the pressure is large and the total effective pressure dips below the vapour pressure of the fluid. At this point, bubbles of the gaseous phase are formed and the hydrodynamics description fails. 
Cavitation is a well studied phenomena that occurs, for example, in pumps and propellers where the induced shockwaves are usually harmful. It has numerous technological application ranging from shock wave lithotripsy (a kidney stone treatment) to ultrasonic cleaning baths. 

While cavitation is fairly well understood, it is not described by hydrodynamics which is based on an expansion in gradient of velocity. When the viscous terms are of the same size as the perfect fluid terms we presumably lose control of the perturbative expansion in gradients. 
In the context of RHIC, as the plasma expands it cools until a crossover occurs at a critical temperature which we take to be $T_c = 190\; \rm{MeV}$.  Finding that a fluid cavitates is therefore the same as figuring out whether the hydrodynamics description breaks down before we reach the critical temperature\footnote{Assuming cavitation does occur, it could have important phenomenological implications on photon and dilepton production \cite{Bhatt:2011kx}.}.

This question has been looked into (with RHIC intial data in mind) using a boost invariant longitudinal flow ($1+1$ dimensional) in \cite{Torrieri:2008ip, Fries:2008ts,rt} (see \cite{Madsen:2009tb} for an analysis in the context of neutron stars) . The analysis was extended to include transverse flow in a $2+1$ dimensional code by Song and Heinz \cite{song}. 
 In this paper, we will follow the (1+1) dimensional analysis of  Rajagopal and Tripuraneni (RT) \cite{rt}. Their approach uses a mixture of inputs from lattice simulation of QCD as well as results borrowed from the conformal plasma:
\begin{enumerate}
\item 
They use a phenomenological model for the equation of state fitted to lattice data and they assume a vanishing chemical potential.
\item They assume a phenomenological bulk viscosity (again from lattice calculations) which rises sharply near the crossover temperature while the shear viscosity to entropy ratio is assumed to be a few factors within the leading order AdS/CFT result of $1/4\pi$.
\item  For the second order transport coefficients (like relaxation times) they take the AdS/CFT values for a {\it conformal} plasma.
\end{enumerate}
They find that the rise in bulk viscosity can lead to cavitation while pointing out that large relaxation times could hinder this effect. 
The analysis of RT  will be the benchmark to which we will compare in this paper and we denote their results $QCD-RT$ in the various figures.
It would be interesting to study cavitation in a theory close to QCD where the AdS/CFT let us calculate and predict all the relevant coefficients.  

 In this paper, we look at cavitation in $\mathcal{N}=2^*$ model whose transport properties have been investigated in some details by Buchel et al. \cite{buchel1} and Buchel and Pagnutti \cite{buchel2}. In particular, we study the 1+1 d boost invariant expansion following Rajagopal and Tripuraneni but with the transport parameters as dictated by this model{\footnote {The numerical data has been kindly supplied to us by Alex Buchel.}.} This model mimics several properties that the sQGP is thought to have. It shows a rising bulk viscosity near the phase transition temperature and a low shear viscosity to entropy density ratio. While these features are in common with \cite{rt}, the relaxation times are not--they rise sharply near the phase transition temperature, quite unlike what happens in the conformal case where they are inversely proportional to temperature. This behaviour, which has been anticipated by Song and Heinz \cite{song}, is responsible for a ``critical slow down". We find that using a normalized data for ${\mathcal N}=2^*$ which sets $T_c\approx 190\;  \rm{MeV}$ and using an initialization time of $\tau_i \approx 0.5 fm/c$, cavitation {\it does not} occur in this stringy example. The main reason for this is the lower value of bulk viscosity and higher value of pressure in these models as compared to the ones used in RT.

We then repeat the analysis of RT using a phenomenological formula for the relaxation time that exhibit critical slow-down as seen in the $\N=2^*$ plasma
\be\label{slowdown}
\tau_\Pi^\zeta = \tau_{\pi}^\eta = \frac{C}{T_c \sqrt{1-T_c/T}}\; .
\ee
The $\N=2^*$ relaxation is well approximated by the formula above (near $T_c$) with $C = 0.08$. We find that cavitation is prevented for values of $C\gsim 2$. We also analyze the situation where the equation of state is taken from the lattice but the bulk viscosity and the relaxation time are taken from the $\N=2^*$ data. We find that cavitation is prevented mainly because of the lower bulk viscosity of the $\N=2^*$ compared to the lattice result. 

Our conclusion is that critical slow-down hinders cavitation but one needs to have a bigger relaxation time than the naive value given by the $\N=2^*$ toy model.  
It is interesting that there is no range of parameters that makes the toy model $\N=2^*$ cavitate before it reaches the phase transition. We give a semi-analytical proof of this behavior.  

This paper is organized as follows. We review the second order hydrodynamics equations in \S II. In \S III, we consider the possibility of cavitation in a conformal plasma and give an analytic argument that this cannot happen. In \S IV, we turn to the numerical study of the 1+1 d boost invariant flow equations using the ${\mathcal N}=2^*$ model. We conclude with a discussion in \S V. The basic assumptions used in RT are reviewed in an appendix.

\section{Boost Invariant Viscous Fluid}

We are interested in a relativistic fluid whose stress-energy momentum tensor is 
\be
T^{\mu\nu} = \e u^\mu u^\nu - p \Delta^{\mu\nu} + \Pi^{\mu\nu}
\ee
where $\epsilon$ and $p$ are the fluid energy density and pressure respectively, $u^\mu$ is the fluid 4-velocity (with $u^\mu u_\mu = 1$), $\Pi^{\mu\nu}$ is the viscosity tensor and $\Delta^{\mu\nu} = g^{\mu\nu} - u^\mu u^\nu$ is the projection tensor orthogonal to the 4-velocity.  The viscosity tensor is also orthogonal to $u^\mu$ with $u_\mu \Pi^{\mu\nu} = 0 $ and $g^{\mu\nu}$ is the metric with signature   $(+, -,-,-)$.  The equation of motion of hydrodynamics is just the conservation of stress-energy
\be
\nabla_\mu T^{\mu\nu} = 0
\ee
where $\nabla_\mu$ is the geometric covariant derivative. We introduce the following notation 
\bea
D  & \equiv & u^\mu \nabla_\mu\; ,\\
 A^{<\mu\nu>}&  = & \left((\De^{\mu\a}\De^{\nu\b} + \De^{\mu\b}\De^{\nu\a}) -\frac23\De^{\mu\nu}\De^{\a\b} \right) A_{\a\b} \; .\nonumber
\eea
We will be considering a purely longitudinal flow. The conservation of stress-energy in the longitudinal direction $u_\mu T^{\mu\nu} = 0$ is 
\be\label{eom}
D\epsilon + (\epsilon +p) \nabla \cdot u + \Pi^{\mu\nu}\nabla_\mu u_\nu =0\; .
\ee
The most general viscous tensor at second order in perturbation theory has been worked out in \cite{Romatschke:2009kr}. Working in flat space with no vorticity reduces the number of terms. Writing $\Pi^{\mu\nu} = \pi^{\mu\nu} + \De^{\mu\nu} \Pi$ we find
\bea
\pi^{\mu\nu}&  = &\eta \nabla^{<\mu}u^{\nu>} - \eta\te \left(D (\nabla^{<\mu}u^{\nu>} )+ \frac{\nabla \cdot u}{3} \nabla^{<\mu}u^{\nu>} \right)\nonumber\\
&&- \lambda_1 \nabla^{<\mu}u_{\lambda}\nabla^{\nu>}u^{\lambda} \label{pino}
\eea
while the trace is given by 
\be\label{Pino}
\Pi = - \zeta \nabla\cdot u + \zeta\tz D(\nabla\cdot u)
\ee
where in the last expression we neglected two second order corrections which are not zero even in our purely longitudinal flow (the coefficient $\xi_{1,2}$ in \cite{Romatschke:2009kr} are set to 0). This is a common procedure. 
The coefficients in front of each viscous terms must be determined experimentally or via some theoretical model. They can depend on temperature or time and they are thus general scalar functions. They are the shear viscosity $\eta$, the bulk viscosity $\zeta$, the relaxation times $\te,\tz$ and $\la_1$. 

At this point this second order hydrodynamic (just like the first order) is acausal. The viscous tensor responds instantaneously to change in the local fluid properties. It has been shown that the superluminal modes are not fundamentally a problem since they corresponds to modes outside of the regime of the effective theory of 
hydrodynamics \cite{Geroch:1995bx, Geroch:2001xs, Kostadt:2000ty}. Nevertheless they pose problems when 
one tries to solve the system of equation numerically. Mueller, Israel and Stewart \cite{Mueller1, Israel:1976tn, IS0b, Israel:1979wp} have proposed a solution which consist of extending the set of variables to include $\Pi$ and $\pi^{\mu\nu}$ and transforming their constitutive relations Eq.~\eqref{pino} and \eqref{Pino} into differential equations.

To do this, we adopt the approach of \cite{Baier:2007ix} (see also \cite{Bhattacharyya:2008jc}) where we simply put the first order solution in the second order terms. The first order solutions are  $\pi^{\mu\nu} = \eta \nabla^{<\mu}u^{\nu>}$ and $\Pi= \zeta \nabla\cdot u$ and substituting in the second order terms,
\bea \label{constitutive}
\pi^{\mu\nu} &= &\eta \nabla^{<\mu}u^{\nu>} - \te \left(\left<D\pi^{\mu\nu}\right>+ \frac{4\nabla \cdot u}{3} \pi^{\mu\nu}\right)\nonumber\\
&&- \frac{\lambda_1}{\eta^2} \pi^{<\mu}_\la\pi^{\nu>\la}\nonumber\; ,\\
\Pi &=& -\zeta \nabla\cdot u - \tz D\Pi
\eea
where we have used the relation $D\eta = -\eta\nabla\cdot u$. This follows from assuming that $\eta \propto s$ where $s$ is the entropy density which obeys $D\ln s = -\nabla\cdot u$ \cite{Romatschke:2009kr, Baier:2007ix}. 
In the second equation, we have dropped the term $ - \tz \nabla\cdot u D\zeta$ which we checked had a very small effect on our analysis. Eq.~\eqref{eom} and \eqref{constitutive} now form an hyperbolic set of equations which is causal and well posed for numerical methods. 

As we already mentioned, we are interested in a boost invariant longitudinal flow in the z direction. Instead of Minkowski coordinate $(t,z)$, it is convenient to change to $(\tau,\xi)$ where $\tau \equiv \sqrt{t^2-z^2}$ is the proper time and $\xi =  {\rm arctanh}(z/t)$. In these coordinates the metric is $g_{\mu\nu} = {\rm diag}( 1, -1,-1, -\tau^2)$, $u^\mu = (1,0,0,0)$ and boost invariance implies that all quantities are independent of $\xi$. Assuming boost invariance and longitudinal flow simplifies the stress-energy tensor to
\bea
T^{\mu\nu}& =& \left(
\begin{array}{cccc} 
\epsilon & 0 & 0 & 0\\
0 & p & 0 & 0\\
0 & 0 &  p& 0\\
0 & 0 & 0 & p/\tau^2 \end{array}\right)+ \nonumber\\
&&
 \left(
\begin{array}{cccc} 
0 & 0 & 0 & 0\\
0 & \Pi+\frac12\Phi & 0 & 0\\
0 & 0 &   \Pi+\frac12\Phi & 0\\
0 & 0 & 0 &  (\Pi-\Phi)/ \tau^2 \end{array}\right)\; .
\eea
It will be useful to define the transverse and longitudinal effective pressure as
\bea
P_{\perp} & \equiv & p +\Pi + \frac12\Phi\; ,\\
P_\xi & \equiv & p+\Pi -\Phi\; .
\eea
In this coordinate system, the evolution equations are
\bea
\frac{\partial \epsilon}{\partial \tau}&=& -\frac{\epsilon+p+\Pi-\Phi}{\tau}\,,\label{energy}\\
\tau_\Pi^\eta \frac{\partial \Phi}{\partial \tau}&=& \frac{4\eta}{3\tau}-\Phi-\frac{4\tau_\Pi^\eta}{3\tau}\Phi - \frac{\la_1}{2\eta^2} \Phi^2\label{shear}\; ,\\
\tz \frac{\partial \Pi}{\partial \tau}&=& -\frac{\zeta}{\tau} -\Pi\; .\label{bulk}
\eea

\section{Cavitation in a conformal plasma}

To get an idea of what we are going after, let us start with a conformal plasma ($\zeta = 0$ and $p =\epsilon/3$). Consider the situation where the hypothetical cavitation occurs at time $\tau=\tau_f$ corresponding to temperature $T=T_f$ and substitute
\begin{equation*}
p=\epsilon/3\,,\quad \epsilon=\epsilon_0 T^4\,, \quad \frac{\eta}{s}=\frac{b}{4\pi}\,,\quad \tau_\pi^\eta=\frac{c}{T}\,,\quad \lambda_1=\frac{ds}{T}\,,
\end{equation*}
in Eq.~(\ref{energy}) and  (\ref{shear}) where $b,c,d$ are unspecified numerical coefficient and $s$ is the entropy density. 
Consider a series solution for $T$ and $\Phi$ around $\tau=\tau_f$. This is given by
\be  
T=T_f-\frac{T_f}{4\tau_f}(\tau-\tau_f)+\cdots\,,\quad \Phi=\frac{\epsilon_0}{3} T_f^4+\phi (\tau-\tau_f)+\cdots\,,
\ee
with
\be
\phi=-\frac{T_f^4\epsilon_0}{9 b^2 c \pi^2 \tau_f}[b^2\pi^2(4 c+3 T_f\tau_f)+6 d \pi^4 T_f \tau_f-4 b^3 \pi)]\,.
\ee
The longitudinal pressure $p-\Phi$
\bea
p-\Phi & = & (\tau-\tau_f) \frac{\epsilon_0 T^4}{9 b^2 c \pi^2 \tau_f}\left[-4 b^3 \pi\right.\nonumber\\
&& \left.+b^2 \pi^2(c+3 T_f \tau_f)+6 d \pi^4 T_f \tau_f\right] \,,
\eea
so that in order to get cavitation we need for $\tau>\tau_f$, $p-\Phi < 0$. This translates into
\be
b^2 \pi  (c\pi -4 b+3 T_f \tau_f \pi)+6 d \pi^4 T_f \tau_f < 0\,.
\ee
From this we can conclude the following.
\begin{itemize}
\item Increasing $\eta/s$ aids in cavitation.
\item Increasing $\tau_\pi^\eta$ hinders cavitation.
\item Increasing $\lambda_1$ hinders cavitation.
\end{itemize}
Now the last term proportional to $d$ is positive so it is a necessary condition that the term proportional to $b$ must be negative for cavitation to occur. There will not be any cavitation unless
\be
r \equiv \frac{1}{T_f\tau_f} > \frac{3\pi}{4b-\pi c}\; .
\ee

Now imagine that after $\tau=\tau_i$ the system has a hydrodynamic evolution until cavitation occur at $\tau=\tau_f$. The cavitation temperature should be greater than the phase transition (crossover in the QCD plasma case) temperature $T_f >T_c$, 
\be
r = \frac{1}{T_f\tau_f} <\frac{1}{T_c \tau_i} \,.
\ee
Now at RHIC, $T_c\sim 200 \MeV$, $\tau_i \sim 0.5 fm/c = 1/400 \MeV$ such that the bound roughly translates into
\be
r<2\,.
\ee
Plugging in the conformal values $b=1$ and $c = \frac{2-\log(2)}{2\pi}$, we have $r\sim 2.8$ so we conclude that this condition cannot be met and cavitation does not occur. In fact since the term proportional to $d$ is positive and comparable to the rest of the terms, its presence only makes cavitation harder.

We now turn to the analysis in QCD and $\N=2^*$ where we know the bulk viscosity is not zero and cavitation is more
likely to happen.
\section{Non-Conformal Case}
We compare the behavior of $\N=2^*$ to the results of RT (which we denote $QCD-RT$ in the graphs). Let us summarize what they did (see appendix \ref{RT} for details). They took a phenomenological approach where all parameters have their conformal values with the exception of the bulk viscosity and the equation of state. They then scan the parameter space. Clearly the presence of bulk viscosity enhances the chances of cavitation since at first order $\Pi = -\zeta/\tau$ and $P_\xi \propto \Pi$.  The relaxation time on the other hand tends to slow the rise of $\Pi$ and it therefore hinders cavitation.

They find that cavitation indeed sets in before $T_c$. The weakest link in this analysis is the choice of the relaxation time which was imported from the conformal plasma calculated using AdS/CFT. There exists a well known AdS/CFT model studied extensively by Buchel and collaborators called the ${\mathcal N}=2^*$ model which has a bulk viscosity \cite{buchel1, buchel2}. It is interesting to ask whether this model exhibits any sign of cavitation along the lines of Rajagopal and Tripuraneni's analysis. Let us first review some salient features of this model.

\subsection{$\N=2^*$}
\begin{figure} [ht]
\includegraphics[width=0.5\textwidth,angle=0]{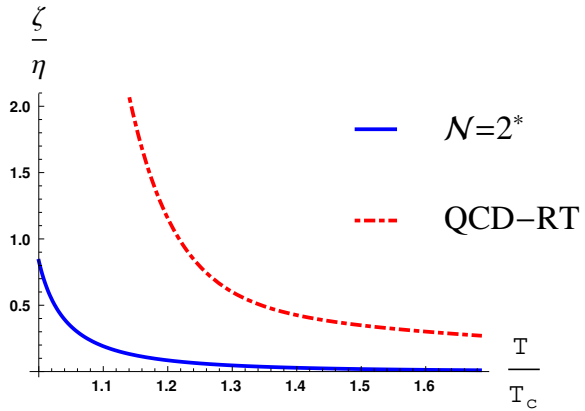}
\caption{Comparison between the ${\mathcal N}=2^*$ bulk viscosity and the phenomenological bulk viscosity fitted to lattice QCD calculation and used in RT}\label{zetabyeta}
\end{figure}

${\mathcal N}=2^*$ is a mass deformation of ${\mathcal N}=4$ and is dual to the Pilch-Warner geometry in IIB. Typically one needs to resort to numerics as the resulting equations are too hard to solve exactly. On the gravity side, there are two scalar fields in a certain potential. One can give masses $m_b, m_f$ to bosonic and fermionic components of the ${\mathcal N}=2$ hypermultiplet respectively. We will consider $m_f=0$ with $m_b\neq 0$. When $m_f^2<m_b^2$ this model exhibits a second order phase transition, in our case when $m_b/T_c=2.33$. We will normalize the free energy such that the phase transition occurs at 190 MeV and choose the number of colour to be $N_c=3$. At finite temperature, the fluid/gravity correspondence analysis of this model yields a bulk viscosity which is thought to respect the Buchel bound $\zeta/\eta \geq 2 (1/3-c_s^2)$ where $c_s$ is the velocity of sound \cite{Buchel:2007mf}. The bulk viscosity rises up sharply near the phase transition (but not as sharply as in QCD). This is illustrated in Fig.~\ref{zetabyeta}.
\begin{figure} [ht]
\includegraphics[width=0.5\textwidth,angle=0]{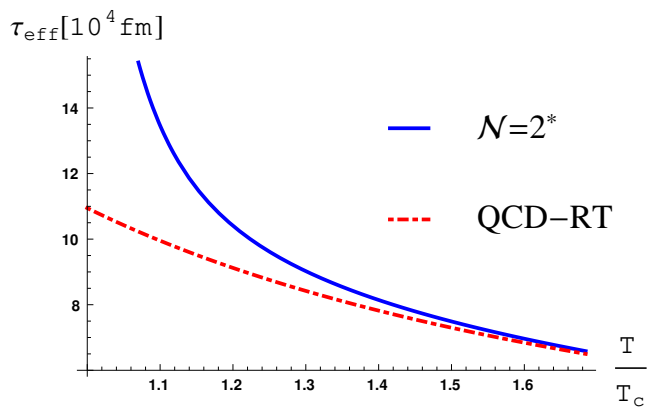}
\caption{Comparison between effective relaxation times. The $\N=2^*$ model shows a sharp rise near $T=T_c$, a sign of critical slow-down.}\label{taus}
\end{figure}

Another crucial difference is the behaviour of the relaxation times as a function of $T/T_c$. The $\N=2^*$ model predicts the following effective relaxation time
\be
\tau_{eff} = \frac{\tau_\Pi^\eta + \frac34\frac{\zeta}{\eta}\tau_\Pi^\zeta}{1+\frac34\frac{\zeta}{\eta}}\; .
\ee
which exhibits a peak 
\be
\tau_{eff} T_c \propto |1-T_c/T|^{-1/2}
\ee
near $T= T_c$.  Note that from the gravity side of the $\N=2^*$, we can only obtain this particular combination of the two relaxation times. We parametrize our ignorance of the repartition between the two times with the parameter $\gamma$,
\begin{align}
\tau_\pi^\eta & = \frac{1+\frac34\frac{\zeta}{\eta} }{1+\gamma \frac34 \frac{\zeta}{\eta}} \tau_{eff}
& \tau_\Pi^\zeta  = \gamma \frac{1+\frac34\frac{\zeta}{\eta} }{1+\gamma \frac34 \frac{\zeta}{\eta}} \tau_{eff}
\end{align}
 
For $\gamma = 1$, we have that $\tau_\pi^\eta = \tau_\Pi^\zeta$, for $\gamma = \infty$, $\tau_\pi^\eta =0$ and  $\tau_\Pi^\zeta = \tau_{eff}\left(\frac{4\eta}{3\zeta}+1\right)$ while for $\gamma = 0$, we have $\tau_\Pi^\zeta =0$ and  $\tau_\pi^\eta = \tau_{eff}\left(1+ \frac34\frac{\zeta}{\eta}\right)$. In the following we will test these three different values of $\gamma$. 
The behaviour of $\tau_{eff}$ as a function of temperature is illustrated in Fig.~\ref{taus}. As we see, the effective relaxation time rises sharply as $T_c$ is approached.  That this plays an important role was already anticipated in the work of Song and Heinz \cite{song}.  Together with the fact that the bulk viscosity in this model does not become too big near $T_c$ with $\zeta/\eta \approx 0.8$ makes cavitation harder. 

As in RT we choose the following initial conditions $\Phi(\tau_i)=0, \Pi(\tau_i)=0$, $\tau_i=0.5$ fm/c and we take the initial energy density to be $\epsilon = 15.8\; \rm{GeV}/\rm{fm}^3$. This initial energy is a bit uncertain but we find that it does not affect the answer greatly.  For the lattice equation of state Eq.~(\ref{QCDeos}), this initial energy density corresponds to an initial temperature of $T = 305\; \rm{MeV}$. For the $\N=2^*$ equation of state, the initial temperature is lower at $T =245 \;\rm{MeV}$. Given that we stop the evolution at the same temperature in both cases (when $T=190 \; \rm{MeV}$), the evolution of the $\N=2^*$ is shorter. 
 
\begin{figure} [ht]
\begin{center}
\includegraphics[width=0.5\textwidth,angle=0]{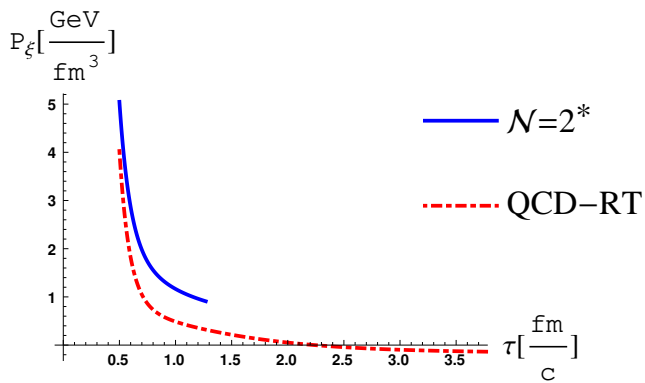}
\caption{The longitudinal pressure for ${\mathcal N}=2^*$ compared with the findings of RT. The string model reaches the critical temperature faster and does not cavitate before that. For the QCD-RT model, cavitation occurs near $\tau \sim 2.5$ fm/c}\label{comparison}
\end{center}
\end{figure}

\begin{figure} [ht]
\begin{center}
\includegraphics[width=0.5\textwidth,angle=0]{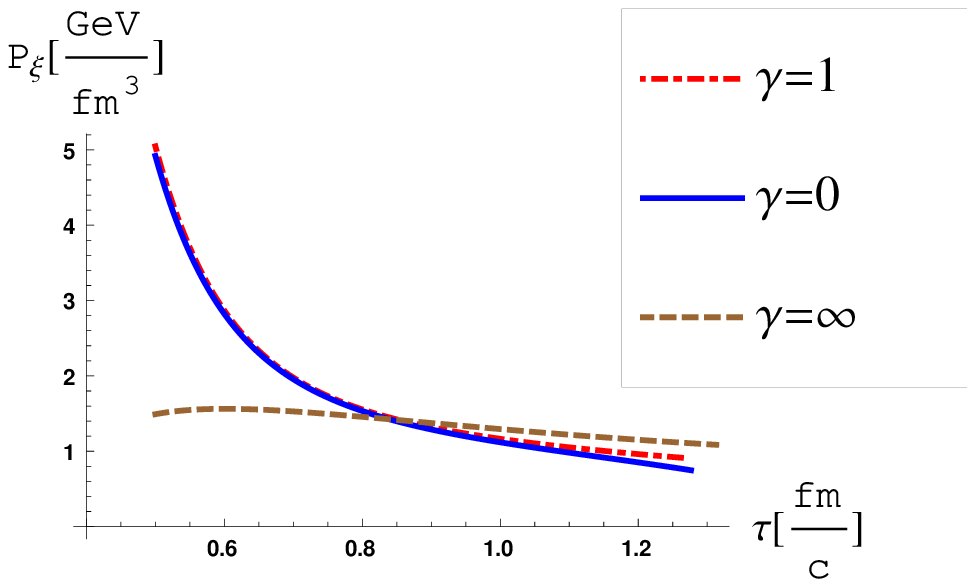}
\caption{The longitudinal pressure for ${\mathcal N}=2^*$ for different values of the parameter $\gamma$. }\label{gamma}
\end{center}
\end{figure}

In Fig.~\ref{comparison}, we plot the longitudinal pressure for the $\N=2^*$ plasma compared to RT. 
The longitudinal pressure decreases with time but remains positive all the way up to $T_c$. We have checked that the reason for this is primarily the smaller values of bulk viscosity in this model. In fact substituting the ${\mathcal N}=2^*$ relaxation times into the analysis of RT still leads to cavitation in their model.
In Fig.~\ref{gamma}, we show how the longitudinal pressure is affected by varying $\gamma$ between the three limiting values of $0,1,\infty$ for the $\N=2^*$ model. For $\gamma = \infty$, the behavior is quite different but the system still does not cavitate.

\subsection{No cavitation for $\N=2^*$}
\begin{figure} [ht]
\begin{center}
\includegraphics[width=0.5\textwidth,angle=0]{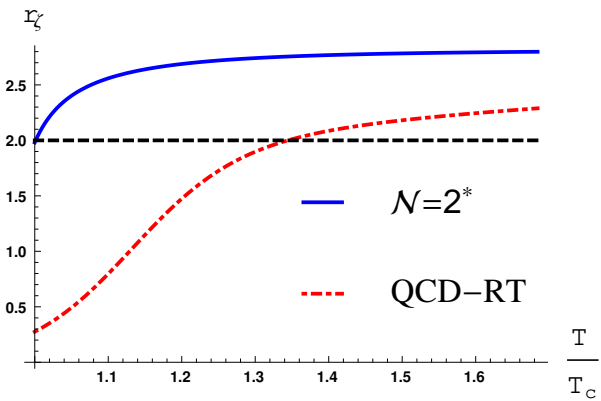}
\caption{$r_\zeta$ as a function of temperature. Cavitation occurs when $r_\zeta <2$. 
}\label{rzeta2}
\end{center}
\end{figure}

As in the conformal case where a series solution enabled us to derive an inequality on the parameters, with some reasonable approximations, we will be able to derive a simple inequality for the non-conformal $\N=2^*$ plasma for cavitation to occur. Let us assume that cavitation occurs at time $\tau=\tau_f$ such that $T>T_c$.  We will further assume that the plasma is close to conformality or in other words $p\sim \epsilon/3$ which is accurate within 10\% error for the $\N=2^*$ plasma but is substantially different for the QCD plasma. For the QCD plasma considered in RT, the pressure works out to be smaller compared to the energy and it turns out that it will make the inequality only stronger (in other words cavitation occurs for the QCD plasma of \cite{rt} at an earlier time than what will be predicted by this analysis).
We begin by subsitituting
\begin{equation*}
p=\epsilon/3\,,\quad \epsilon=\epsilon_0 T^4\,, \quad \frac{\eta}{s}=\frac{b}{4\pi}\,,\quad \tau_\Pi^\eta= \tau_\Pi^\zeta\; .
\end{equation*}
We take $\lambda_1 = 0$ and $\gamma =1$ for simplicity and
\begin{equation*}
\tz=\tz_0+\tz_1(\tau-\tau_f)\,,\quad \zeta=\zeta_0+\zeta_1(\tau-\tau_f)\,,
\end{equation*}
in Eq.~(\ref{energy}) ,  (\ref{shear}) and (\ref{bulk}). Consider a series solution for $T$ , $\Phi$ and $\Pi$ around $\tau=\tau_f$. This is given by
\be  
T=T_f-\frac{T_f (4 \epsilon_0-\epsilon_1)}{12\epsilon_0 \tau_f}(\tau-\tau_f)+\cdots\,, \Phi=\frac{\epsilon_1}{3} T_f^4+\phi (\tau-\tau_f)+\cdots\,,
\ee
and
\be
\Pi=(\epsilon_1-\epsilon_0)\frac{T_f^4}{3}+\pi_1(\tau-\tau_f)+\cdots\,.
\ee
Here we have parametrized the deviation from conformality by $\epsilon_1-\epsilon_0$. Now we solve for $\phi$ and $\pi_1$ and we find that the longitudinal pressure around $\epsilon_0\approx \epsilon_1$ is
\be
P_\xi=\frac{\left[-4b \epsilon_0T_f^3+  \pi ( 3\tau_0 T_f^4 \epsilon_0 -9\zeta_0 + T_f^4\tz_0 \epsilon_0)\right]}{9\pi \tau_0\tz_0}(\tau-\tau_f)\,.
\ee
Now the analysis is exactly as in the conformal case. In this case defining (with $\epsilon_f=\epsilon|_{T=T_f}$)  
\be\label{rzetaeq}
r_\zeta \equiv \frac{1}{T_f\tau_f} > \frac{3\pi \epsilon_f}{4 b \epsilon_f+9\zeta_0\pi T_f -\epsilon_f \pi \tz_0 T_f}
\ee
and choosing the RHIC initial time and temperature as in the conformal case leads to cavitation happening only if $r_\zeta<2$.  Some comments are in order regarding the choice of $\epsilon_1 \approx \epsilon_0$.  
Our numerical analysis show that the $\N=2^*$ plasma has $\Pi\rightarrow 0$  at late time corresponding to $\epsilon_1= \epsilon_0$.  This is both because of lower bulk viscosity and because of large relaxation time. 
 In RT, on the other hand we have $\Pi \approx - \Phi$ at late time (so $\epsilon_1 \approx \epsilon_0/2$). So our current approximation is only good for the $\N=2^*$ plasma. 
 As we see in Fig.~\ref{rzeta2}, cavitation does not happen for $\N=2^*$ while RT on the other hand clearly cavitates. Since the QCD plasma of \cite{rt} is far from conformality, our approximation is rather crude and predicts cavitation to occur when $T$ is between $1.3 T_c$ and $1.4 T_c$ while the numerics gives around $1.7 T_c$. The reasons for this discrepancy is that the QCD pressure is lower than the conformal approximation and  $\epsilon_1 \approx \epsilon_0$ at late time.  Accounting for both effects lead to an earlier cavitation time for RT.

\subsection{Critical slow-down}
\begin{figure} [ht]
\begin{center}
\includegraphics[width=0.5\textwidth,angle=0]{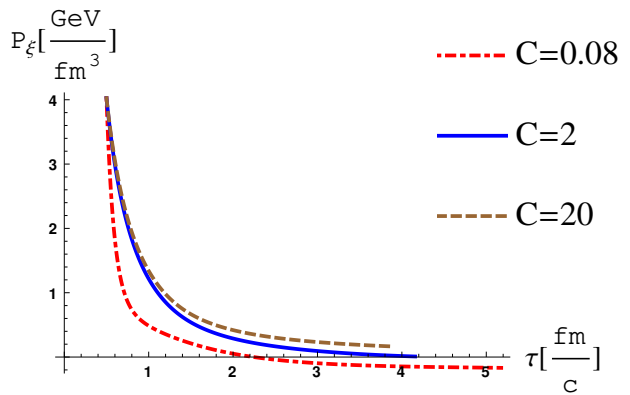} 
\caption{This is the longitudinal pressure for a plasma with equation of state Eq.~(\protect \ref{QCDeos}) and (\protect \ref{zetabys}) but with a phenomenological relaxation time that exhibit critical slow-down for different value of $C$. }\label{QCDpheno}
\end{center}
\end{figure}

It is interesting to ask if the critical slow down phenomena where the relaxation time shoots up near $T_c$ can inhibit cavitation in the QCD plasma considered in RT.  We take the equation of state and bulk viscosity from the lattice simulation as in Eq.~(\ref{QCDeos}) and (\ref{zetabys}) but instead consider a phenomenological relaxation time of the form \cite{buchel1}
\be
\tau_\Pi^\zeta = \tau_{\pi}^\eta = \frac{C}{T_c \sqrt{1-T_c/T}}
\ee
For $C = 0.08$ we find that this is a good fit to the $\N=2^*$ data for a large range of temperature although it is important to mention that the relaxation time does not actually diverge in the $\N=2^*$ plasma at $T= T_c$. As can be seen from Fig.~\ref{QCDpheno}, cavitation is prevented when one reaches value of $C\gsim 2$. We conclude that critical slow-down does prevent cavitation but one needs a bigger effect than what is seen in the $\mathcal{N}=2^*$ plasma. 

\section{Conclusion}

In this paper we investigated if the ${\mathcal N}=2^*$ model in string theory exhibits cavitation and we contrasted our finding with the analysis in QCD following closely \cite{rt} (RT). We found that for this choice of normalizations and initialization conditions this model does not cavitate before $T_c$ is reached. Primarily, the smaller values of bulk viscosity and secondarily, the sharp rise in the relaxation time near $T_c$  are responsible for this behaviour. It is also likely that quantum effects  as in \cite{qeff} which correct $\eta/s$ and $\zeta$ will play an important role in these studies.

If we made the initial time $\tau_i$ of hydrodynamic evolution smaller, the chance of cavitation becomes higher. For example, for the $\N =4$ plasma, our analysis shows that cavitation could occur for an initialization time less then $0.3$ fm/c while the $\N = 2^*$ plasma could cavitate for $\tau_i < 0.14$ fm/c. Simulations at RHIC suggest $\tau_i$ closer to $1$ fm/c in which case cavitation would not happen. 

As stated in the introduction in order for the longitudinal pressure to become negative, a second order contribution to the pressure has to be as big as a zeroth order contribution. As such, mathematically the gradient expansion approximation used becomes incorrect.   The final value of $P_\xi$ in our simulations (see Fig.~\ref{comparison}) was around 20\% of the initial value indicating that the numerical approximations starts to become suspect. This raises the question if the $P_\xi<0$ condition is a good marker for the breakdown of the hydrodynamics equations. One can perhaps answer this question more accurately using AdS/CFT where a breakdown of the hydrodynamics equations is signalled by the existence of some instability in the gravitational theory. 

\begin{acknowledgements}
We are grateful to Alex Buchel and Rob Myers for useful discussions. We are especially grateful to Alex Buchel for sharing his numerical data which enabled us to extract the bulk viscosity and relaxation time behaviours as functions of the temperature and time; this work would not have been possible without his encouragement and advice. We thank Giorgio Torrieri for useful communication and clarifications. Research at the Perimeter Institute is supported in part by the Government of Canada through NSERC and by the Province of Ontario through the Ministry of Research and Information (MRI). AS acknowledges partial financial support through a Ramanujan fellowship by the Department of Science and Technology, Govt. of India. AK was a summer undergraduate student in the Perimeter Institute.

\end{acknowledgements}

\appendix

\section{QCD equation of state}
\label{RT}
Here we summarize the important assumptions in RT \cite{rt} that we used as a baseline for comparison throughout our analysis. The equation of state is taken from the lattice calculation of \cite{Bazavov:2009zn} and it can be nicely parametrized as follows
\be\label{QCDeos}
\frac{\epsilon-3p}{T^4} = \left(1- \frac{1}{\left(1+\rm{exp}(\frac{T-c_1}{c_2})\right)^2} \right) \left(\frac{d_2}{T^2} + \frac{d_4}{T^4}\right)
\ee
with $d_2 = 0.24 \gev^2$, $d_4 = 0.0054 \gev^4$, $c_1 = 0.2073 \gev$ and $c_2 = 0.0172 \gev$. The crossover temperature is chosen to be $T_c = 190 \mev$. We also have
\be
\frac{P(T)}{T^4} = \int_{T_0}^T dT' \frac{\epsilon - 3p}{T'^5}\; .
\ee
For the various transport coefficients, RT choose
\begin{equation*}
\frac{\eta}{s}=\frac{1}{4\pi}\,,\quad \tau_\pi^\eta=\frac{2-\log 2}{2\pi T}\,,\quad \lambda_1=\frac{s}{8\pi^2T}\,.
\end{equation*}
The bulk viscosity is taken from the lattice \cite{Meyer:2009jp}
\be
\frac{\zeta}{s} = a_1 {\rm exp} \left(\frac{T_c-T}{\Delta T}\right) + a_2 \left(\frac{T_c}{T}\right)^2\label{zetabys}
\ee
with $a_1 = 0.901$, $a_2= 0.061$ and $\Delta T = \frac{T_c}{14.5}$.
RT varied their parameters and found the following:
\begin{itemize}
\item Cavitation is insensitive to changes in $\tau_\pi^\eta$ and $\lambda_1$ 
\item Cavitation disappears when $\tau_\Pi^\zeta$ is increased by 40 from its baseline value taken to be $\tau_\Pi^\zeta=\tau_\pi^\eta$.
\end{itemize}

The latter is in agreement with our finding that cavitation does not occur if we allow for a critical slow-down of the form Eq.~\ref{slowdown} with $ C\gsim 2$.

\end{document}